# Modified Trial Division Algorithm Using KNJ-Factorization Method To Factorize RSA Public Key Encryption

Nidhi Lal[1], Anurag Prakash Singh[2], Shishupal Kumar[3]
Wireless Communication and Computing
Indian institute of Information Technology
Allahabad, India
Nidhi.2592@gmail.com[1]
Anuragvns05@gmail.com[2]
skystargreen2@gmail.com[3]

**Abstract-** The security of RSA algorithm depends upon the positive integer N, which is the multiple of two precise large prime numbers. Factorization of such great numbers is a problematic process. There are many algorithms has been implemented in the past years. The offered KNJ -Factorization algorithm contributes a deterministic way to factorize RSA N=p*q. The algorithm limits the search by only considering the prime values. Subsequently prime numbers are odd numbers (apart from 2) accordingly it also requires smaller number steps to factorize RSA. In this paper, the anticipated algorithm is very simple besides it is very easy to understand and implement. The main concept of this KNJ-factorization algorithm is, to check only those factors which are odd and prime. The proposed KNJ- Factorization algorithm works very efficiently on those factors; which are adjoining and close to √N. The proposed factorization method can speed up if we can reduce the time for primality testing. It fundamentally decreases the time complexity.

## 1. INTRODUCTION

Factorization is a reverse process of multiplication. It is the act of splitting an integer into a set of smaller integers (factors) which, when multiplied collected, form the original integer so it is a hard process to find the factors of very large numbers. It has not been demonstrated that factoring requirement is difficult, and there residues a chance that a rapid and easy factoring method might be exposed [6].

The private key is period coupled and it is mathematically related to the corresponding public key. Hence, it is repetitively probable to attack a public-key system by originating the private key commencing the public key. For occurrence, specific Public-key cryptosystems are considered such that deriving the private key from the public key involves the attacker to factor a large number, therefore, it is computationally infeasible to implement the derivation. This is principally the significant idea of the RSA public-key cryptosystem [5].

A deterministic factoring algorithm based on mathematical ideas of long multiplication was implemented to limit the possible values of p and q. Since the proposed algorithm is sequential in nature, so it requires more steps to find different combinations of p and q; is useful for a smaller amount storage [1]. Alternative algorithm to break RSA based on Fermat Factorization method was implemented. It is simple and accessible. Despite the fact that, it works optimum when there is a factor in close proximity the square-root of N [2]. A new straightforward factorization algorithm base on Trial Division method was implemented. It uses awfully easy arithmetic operations. But it takes more time to check all possible odd numbers closest to square root of N [4]. The main idea of our proposed KNJ-Factorization algorithm is to find out the values of p and q, i.e. the factors of N. Let's see how to split the RSA is probable when only public key couple (N, e) is known. To decrypt a cipher one necessitate to discover d and $d = e^{-1} \mod(\varphi)$, here $\varphi = (p-1)*(q-1)$. If one can find p and q then it is easy to find the value of d. So one have to factorize N, where N=p*q. accordingly factoring an RSA N=p*q would allow an assailant to form out the private key. Therefore, any person who can factor the N, can decrypt message. The protection and security of cryptosystem of the RSA is built upon the procedure of factorization of great integers [8].

In this paper, a modified Trial Division Method based on Trial Division method is proposed. By way of using this algorithm; we can factorize all positive Integer numbers, which are the simple product of two numbers that have exhibits to the property of prime number.

In RSA, technique of factorization does elevated computation [9]. The proposed KNJ-Factorization algorithm checks only those values which are prime. To check any number to be prime number a Primarily Testing is necessary and since prime numbers are odd (excluding 2), so we can also reduce the search time as considering only odd values. JAVA background platform is used for numerous analyses.

## 2. RELATED WORK

An algorithm for attacking RSA scheme based on the knowing public key (e, n) work efficiently if the decryption key d is small [2].

In [3], proposed an innovative procedure to break the RSA. This algorithm divide Fermat Factorization method in two part first is one is, factorize number with respect floor function of square root of N, to get maximum factors that are neighbour to the √(N), second is if don't get positive integer value of square root (square root of N), then sequence between floor(√(N)) to N.

An innovative technique has been introduce [4], to factorize RSA modulus N. This was established on Trial Division method and customs simple arithmetic operations for finding the factors which are nearby to √(N).

## 3. RSA APPROACH

RSA is a public key cryptographic algorithm. It is established on assumed difficulty of factoring enormous numbers i.e. factoring problem. It is the principal algorithm acknowledged to be appropriate signing as well as encryption, and was one of the great developments in public key cryptography.

In public key cryptography scheme, there are basically two number of keys, those names are public and private key which are generated by means of the RSA algorithm and the encryption procedure in the RSA mechanism uses the public key whereas decryption process uses the private key. Neither key will do both function. The Algorithm is as follows [6]:

**Generation of key:**

To create the keys, sender A needed do the following:
1. Generate two large random prime numbers, $p$ and $q$, of nearly identical size.
2. compute $N = p*q$;
3. Compute $(\varphi)$ phi $= (p-1)*(q-1)$.
4. Take an integer $e$, $1 < e < \varphi$, such that $GCD(e, \varphi) = 1$.
5. Calculate the secret exponent $d$, $1 < d < \varphi$, such that $e*d \equiv 1 \pmod{\varphi}$.
6. The public key is $(N, e)$ and the private key is $(N, d)$. Preserve all the values $d$, $p$, $q$ and $\varphi$ secret.

- $N$ is known as the *modulus*.
- $e$ is identified as the public exponent or encryption exponent or just the exponent.
- $d$ is recognized by way of the secret exponent or decryption exponent.

Fig 1: Procedural phases for generation of key

**Procedure for Public key encryption and Decryption:**

**Encryption:**

Sender A encrypts a message $M$ intended for recipient B which decrypt. The correspondent A does the following:
1. Achieves the receiver B's public key $(N, e)$.
2. Represent the plaintext message as a positive integer $M$, where $0 < M < N$.
3. Computes the cipher text $C = M^e \bmod n$.
4. Send the cipher text $C$ to recipient B.

**Decryption:**

To extract the message $M$ from the cipher text $C$, the recipient B does the following:
1. Obtain the cipher text from sender A.
2. Uses his private key $(N, d)$ to compute $M = C^d \bmod N$.

Fig 2: Procedural steps for Encryption and Decryption

**Trial Division Algorithm**

Trial Division is one of the serenest algorithms which does factors of positive integer. It assured to find a factor of N, if the out coming factor is one, then only it further checks all conceivable prime factors of N.

In this algorithm, we only need to test out up to the square root of n, [11-14] meanwhile if suppose variable p is a factor, so q=N/p is also an alternative factor. If number of factors from 2… floor (√ N) (inclusive) are originate, the number is a prime. If N has non trivial factors p, q with p*q=N and p ≤ q, then q≤ (√N).The procedure of the Trial Division factorization is describe further down [7]:

Input: N
Output: Factors of N
1. for p from 2 to floor (√N)
2.     if p divides N then
3.         return p, N/p
4.     end if
5. end for

## 4. PROPOSED METHOD

The proposed KNJ-Factorization scheme gives an algorithm and by use this we can efficiently factorize RSA. This algorithm is not probabilistic. In this algorithm, N is the product of two prime numbers and to factorize this number N, the procedure checks only those numbers which are odd and as well as prime. The steps involved in the KNJ-factorization algorithm are as follows:

1. Let N = P*Q.
2. Compute X=floor (√N).
3. Check If ( X % 2 == 0)
4.     X =X+1.
5. Check If X is Prime
6.     Compute Y=(N/X)
7.     If Y is integer.
8.         Compute P =X and Q =Y and Stop.
9.     Else X = X -2.
10. Else X → X-2, X- 4,…,5, 3.
11. Carry on step 5 to 10, till Y is integer.

**Example 1:**

Let N =55
Compute X=7 (X is prime. Then proceed to step 6)

Compute Y=7.857 (Y is not an integer. Then jump to step 9)
X=5 (X is prime. proceed to step 6) Compute Y=11 (Y is an integer. jump to step 8) Compute P=5, Q=11.

**Example 2:**
Let N =2097491441
Compute X=45798 (Creating X odd)
Compute X=45799 (X is not prime. proceed to step 10) X=X-2,……,X-35826
X=9973 (X is prime. jump to step 6)
Compute Y=210317 (Y is an integer. so proceed to step 8)
 Compute P=9973, Q=210317.

## 5. TESTING AND ANALYSIS

The modified-factorization method, which is called KNJ-Factorization method, gives better results as compared the Trial-Division method.
The KNJ-Factorization method provides efficient results with the minimum number of iterations, hence it reduces the time complexity and increases the speed of the computation. The Trial-Division and the proposed KNJ algorithm was tested on Intel core-i3 PC 2.4 GHz with 3 GB RAM under Microsoft Windows-7 64-bit using JAVA programming. The original and modified Factorization method implementations are shown in Fig. 3.

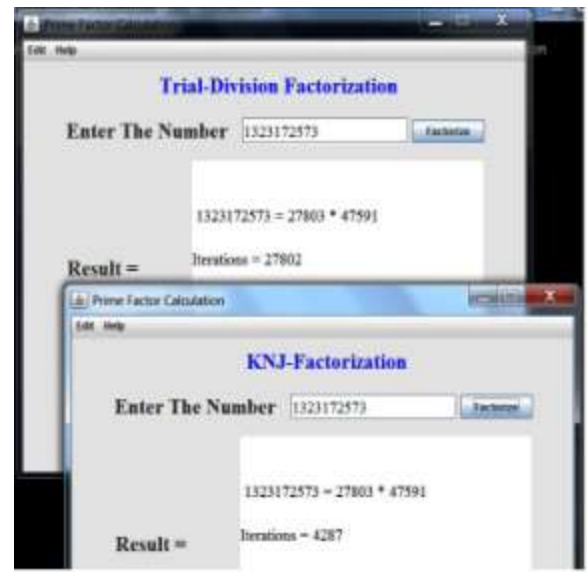

Fig. 3 Implementation and Simulation results of Trial-Division and KNJ-Factorization scheme.

## 6. RESULT

Since, there are many factoring algorithms were developed in the research area of RSA, but we equated and compared some of the results of this algorithm with Trial-Division method. Table 1 indicates particular some of the results and compare among Trial-Division and proposed KNJ-Factorization method.

**Table 1.** Evaluation of KNJ-Factorization with Trial-Division procedure to factorize RSA

| No. of digits in N | N | Iterations (Trial-Division) | Iterations (KNJ Method) |
|---|---|---|---|
| 2 | 55 | 4 | 2 |
| 4 | 1943 | 28 | 9 |
| 6 | 998299 | 822 | 89 |
| 8 | 85928201 | 8752 | 259 |
| 10 | 1323172573 | 27802 | 4287 |

Even the number of digits in N is increases; the proposed algorithm requires less iteration to compute the factors of N as compared to the traditional Trial-Division method. Hence, the total number of iterations in the KNJ-Factorization method is very less as compare to Trial –Division method, therefore KNJ-Factorization method gives better results, increases speed of computation and provides efficient way of factorization.

## 7. CONCLUSION AND FUTURE WORK

In this paper, the proposed algorithm is very efficient and it is very easy to implement using the java platform. The main concept is to check only those factors which are odd as well as those are prime numbers. The proposed KNJ-Factorization algorithm works very efficiently on factors, those are nearby and very closest to √N. The proposed factorization method can give better results and can speed up the computation if we can reduce the time for primality testing. So by this proposed KNJ-Factorization scheme the conclusion comes out that if anyone can factorize the value of N, then there is a possibility to find out the private key very easily and thus the security of RSA will be weak because it can very easily break the security. So, the future work is to find such value of private key which is independent of the value of N.